\documentclass[pra,twocolumn,amsmath,showpacs]{revtex4}
\usepackage{graphicx,bm}
\begin{document}

\hyphenation{mol-e-cul-es}

\author{J. Aldegunde}
\email{E-mail: Jesus.Aldegunde@durham.ac.uk}
\affiliation{Department of Chemistry, Durham University, South
Road, DH1 3LE, United Kingdom}

\author{Ben A. Rivington}
\affiliation{Department of Chemistry, Durham University, South
Road, DH1 3LE, United Kingdom}

\author{Piotr S. \.Zuchowski}
\email{E-mail: Piotr.Zuchowski@durham.ac.uk}
\affiliation{Department of Chemistry, Durham University, South
Road, DH1 3LE, United Kingdom}

\author{Jeremy M. Hutson}
\email{E-mail: J.M.Hutson@durham.ac.uk} \affiliation{Department
of Chemistry, Durham University, South Road, DH1 3LE, United
Kingdom}

\title{The hyperfine energy levels of alkali metal dimers: \\
ground-state polar molecules in electric and magnetic fields}

\begin{abstract}
We investigate the energy levels of heteronuclear alkali metal
dimers in levels correlating with the lowest rotational level
of the ground electronic state, which are important in efforts
to produce ground-state ultracold molecules. We use
density-functional theory to calculate nuclear quadrupole and
magnetic coupling constants for RbK and RbCs and explore the
hyperfine structure in the presence of electric and magnetic
fields. For nonrotating states, the zero-field splittings are
dominated by the electron-mediated part of the nuclear
spin-spin coupling. They are a few kHz for RbK isotopologs and
a few tens of kHz for RbCs isotopologs.
\end{abstract}


\date{\today}

%
\maketitle
\section{Introduction}
\label{intro}

There is great interest in the formation of ultracold molecules
and in achieving molecular Bose-Einstein condensation and Fermi
degeneracy. Molecules can be formed in ultracold atomic gases
either by photoassociation
\cite{Hutson:IRPC:2006,Jones:RMP:2006} or by tuning through
zero-energy Feshbach resonances with magnetic fields
\cite{Hutson:IRPC:2006,Koehler:RMP:2006}. Since alkali metal
atoms are easier to cool than other species, most work on
ultracold molecule formation has focussed on alkali metal
dimers.

There is particular interest in forming ultracold {\em polar}
molecules. Dipole-dipole interactions are both stronger and
longer-range than the quadrupole-quadrupole and dispersion
forces that exist between nonpolar molecules. As a result,
dipolar quantum gases are predicted to have novel properties
\cite{Baranov:2002}. Ultracold dipolar molecules might also be
used in quantum information storage and processing
\cite{DeMille:2002}.

Both photoassociation and Feshbach resonance tuning form
molecules that are initially in highly excited vibrational
states. Quantum gases of such molecules can be formed
\cite{Jochim:Li2BEC:2003,Zwierlein:2003,Greiner:2003}, but they
are long-lived only in very specific cases, such as homonuclear
fermion dimers in the highest vibrational level, tuned to large
scattering lengths \cite{Petrov:2004}. For other cases the
molecules undergo fast inelastic collisions that lead to trap
loss \cite{Herbig:2003,Soldan:2002,Hutson:IRPC:2007}.
Furthermore, even heteronuclear molecules are essentially
nonpolar when they are in weakly bound vibrational states.
Because of this, there is intense current effort directed at
producing ultracold molecules in their absolute ground states,
for which inelastic losses cannot occur and for which
heteronuclear molecules have significant dipole moments. Very
recently, there have been major advances in transferring
Feshbach molecules to deeply bound states by laser-based
methods such as stimulated Raman adiabatic passage (STIRAP)
\cite{Winkler:2007,Ospelkaus:2008,Danzl:2008}. Formation of
quantum gases of ground-state molecules is now within reach.

There has been a considerable amount of work on the energy
levels of homonuclear alkali metal dimers, especially in the
near-dissociation states formed by Feshbach resonance tuning
\cite{Mark:stuck:2007,Mark:spect:2007,Chin:cs2-fesh:2004,Hutson:Cs2:2008}.
However, remarkably little is known about the hyperfine
structure of the energy levels of alkali metal dimers in their
lowest rotational states. The tiny splittings are beyond the
resolution of most spectroscopic techniques. Nevertheless, an
understanding of these energy levels is essential in designing
laser-based methods to produce molecules in specific states and
will be crucial in developing methods to control the resulting
quantum gases. The purpose of the present paper is to
investigate the lowest energy levels of heteronuclear alkali
metal dimers and to explore how they behave in electric and
magnetic fields. We focus here on RbK and RbCs, which are
topical for current experiments.

\section{Theory}
\label{sec:theory}

\subsection{Molecular Hamiltonian}
\label{sec:MolHam}

The Hamiltonian of a diatomic molecule in the presence of
external magnetic and electric fields can be decomposed into
six different contributions: the electronic, vibrational,
rotational, hyperfine, Stark and Zeeman terms. By restricting
our analysis to $^{1}{\rm \Sigma}$ molecules in the ground
electronic state and in a fixed vibrational level, the first
two terms take a constant value and the rotational, hyperfine,
Stark and Zeeman parts of the Hamiltonian can be written
\cite{Ramsey:1952,Brown,Bryce:2003}
\begin{equation}
\label{Htot} H = H_{\rm rot} + H_{\rm hf}  + H_{\rm S}  + H_{\rm Z},
\end{equation}
where
\begin{eqnarray}
 \label{eq:Hrot} H_{\rm rot}  &=& B_v\bm{N}^{2}-D_v\bm{N}^{2}\cdot\bm{N}^{2}; \\
  \nonumber \label{eq:Hhf}H_{\rm hf}  &=& \sum_{i=1}^{2}\bm{V}_{i}:\bm{Q}_{i} \\
&+& \sum_{i=1}^{2} c_{i} \,\bm{N}\cdot\bm{I}_{i}
+c_{3}\,\bm{I}_{1}\cdot\bm{T}\cdot\bm{I}_{2}
+c_{4}\,\bm{I}_{1}\cdot\bm{I}_{2};\\
\label{eq:Hs} H_{\rm S}  &=& -\bm{\mu}\cdot\bm{E}; \\
\label{eq:Hz} H_{\rm Z}  &=& -g_{\rm r}\mu_{\rm N}
\,\bm{N}\cdot\bm{B}
 -\sum_{i=1}^{2}g_{i}\mu_{\rm N}
\,\bm{I}_{i}\cdot\bm{B} (1-\sigma_{i}).
\end{eqnarray}
The three different sources of angular momentum in a $^{1}{\rm
\Sigma}$ diatomic molecule are the rotational angular momentum
$\bm{N}$ and the spins $\bm{I}_{1}$ and $\bm{I}_{2}$ of nuclei
1 and 2. The rotational and centrifugal distortion constants of
the molecule are $B_v$ and $D_v$ (the centrifugal distortion
contribution will not be considered in the calculations). The
hyperfine Hamiltonian of equation \ref{eq:Hhf} consists of four
terms. The first is the electric quadrupole interaction with
coupling constants $(eqQ)_{1}$ and $(eqQ)_{2}$, where $q_{i}$
is the electric field gradient at nucleus $i$ and $eQ_{i}$ is
its nuclear quadrupole moment. The second is the interaction
between the nuclear magnetic moments and the magnetic field
created by the rotation of the molecule, with spin-rotation
coupling constants $c_{1}$ and $c_{2}$. The two remaining terms
represent the tensor and scalar interactions between the
nuclear dipole moments, with spin-spin coupling constants
$c_{3}$ and $c_{4}$ respectively. The tensor $\bm{T}$ describes
the angle-dependence of the direct spin-spin interaction and
the anisotropic part of the indirect spin-spin interaction
\cite{Bryce:2003}.

The Stark and Zeeman Hamiltonians, equations \ref{eq:Hs} and
\ref{eq:Hz}, describe the interaction of the molecule with an
external electric field $\bm{E}$ and magnetic field $\bm{B}$,
where $\bm{\mu}$ is the molecular dipole moment. The Zeeman
Hamiltonian consists of two terms representing the rotational
and nuclear Zeeman effects. The former arises because the
molecular rotation produces a magnetic moment $g_{r}\mu_{\rm
N}\bm{N}$, where $g_{r}$ is the rotational g-factor of the
molecule, which interacts with the external magnetic field. The
latter arises from the interaction of the nuclear magnetic
moments $g_{i}\mu_{\rm N}\bm{I}_i$ with the magnetic field,
where $g_{i}$ is the nuclear g-factor for nucleus $i$ and
$\bm{I}_{i}$ is its nuclear spin. The nuclear shielding tensor
${\bm \sigma}_{i}$ is approximated here by its isotropic part
$\sigma_i$; terms involving the anisotropy of ${\bm
\sigma}_{i}$ are extremely small for the states considered
here. The diamagnetic Zeeman effect is not included in the
Hamiltonian as it causes level splittings less than 1 Hz for
the range of magnetic fields considered in this work.

The nuclear g-factors and quadrupole moments are well known
\cite{Mills:1988}. The dipole moments of KRb and RbCs have been
calculated from relativistic electronic structure calculations
\cite{Kotochigova:2003,Kotochigova:2005}.

\section{Evaluation of the coupling constants}
\label{sec:eval}

Nuclear quadrupole coupling constants have been measured for
several alkali metal dimers as shown in Table \ref{tb:eqqc}.
However, the only such species for which the magnetic coupling
constants have been measured is Na$_2$ \cite{Esbroeck:1985},
and even there the experiments did not resolve hyperfine
splittings for the $N=0$ state. To the best of our knowledge,
no experimental data are available for the hyperfine structure
of the molecules we consider here, KRb and RbCs, in their
ground electronic state. We therefore carry out electronic
structure calculations to estimate them. The electric
quadrupole coupling constants $(eqQ)_{1}$ and $(eqQ)_{2}$, the
nuclear shielding, the spin-rotation constants $c_{1}$ and
$c_{2}$ and the spin-spin coupling constants $c_{3}$ and
$c_{4}$ are evaluated by density-functional theory (DFT) using
the ADF package \cite{ADF1,ADF3}, which uses Slater functions
and allows the inclusion of relativistic corrections. The
rotational g-factor (not implemented in the ADF code) is
evaluated with the DALTON package \cite{Dalton}.

The objective of the present paper is to explore the behaviour
of the molecular energy levels in the presence of external
fields. A detailed discussion of the features and effectiveness
of the many different methods and basis sets available for the
calculation of the coupling constants is beyond the scope of
the work. However, to estimate the reliability of the
functionals and basis sets employed here we compare the
coupling constants obtained for a group of molecules containing
alkali metal atoms with experimental results in tables
\ref{tb:eqqc}, \ref{tb:rc}, \ref{tb:sscc} and \ref{tb:rgf}. For
simplicity we have omitted experimental uncertainties and
vibrational state dependences. It may be seen that the
calculated coupling constants are generally within 30\% of the
experimental values, except in occasional cases where the
experimental values are unusually small (such as $c_4$ for
$^{85}$Rb$^{35}$Cl).

\begin{table} \caption{Comparison of electric quadrupole coupling
constants for alkali metals atoms calculated as described in
the text with experimental values. The units are MHz.}
\label{tb:eqqc}       
\begin{tabular}{llll}
\hline\noalign{\smallskip} Molecule & $(eQq)^{\rm Calc}$ & $(eQq)^{\rm Exp}$ & {\rm Ref.}\\
\noalign{\smallskip}\hline\noalign{\smallskip}
$^{23}$Na$_{2}$ & $-$0.456 & $-$0.459 &  \cite{Esbroeck:1985}\\
$^{39}$K$_{2}$ & $-$0.279 & $-$0.158 &  \cite{Logan:1952}\\
$^{39}$K$^{19}$F & $-$7.87 & $-$7.93 &  \cite{Bonczyk:1967}\\
$^{39}$K$^{7}$Li & $-$0.830& $-$1.03 &  \cite{Dagdigian:1972}\\
$^{39}$K$^{23}$Na & $-$0.671& $-$0.718 &  \cite{Dagdigian:1972} (for K)\\
$^{39}$K$^{23}$Na & $-$0.216& 0.171\footnote[1]
{Only the absolute value was determined experimentally.} &  \cite{Dagdigian:1972} (for Na)\\
$^{85}$Rb$_{2}$ & $-$2.283 & $-$1.1 &  \cite{Logan:1952}\\
$^{85}$Rb$^{19}$F & $-$73.1& $-$70.7 &  \cite{Cederberg:2006}\\
$^{85}$Rb$^{35}$Cl & $-$53.5 & $-$52.8 &  \cite{Cederberg:2006a}\\
$^{85}$Rb$^{79}$Br & $-$46.8 & $-$47.2 &  \cite{Tiemann:1977}\\
$^{85}$Rb$^{127}$I & $-$39.6 & $-$58.9 &  \cite{Tiemann:1976}\\
$^{85}$Rb$^{7}$Li & $-$8.04 & $-$9.12 &  \cite{Dagdigian:1972}\\
$^{133}$Cs$^{19}$F  & 1.30 &  $1.25$ &  \cite{Cederberg:1999}\\
$^{133}$Cs$^{35}$Cl &  1.05 & $\le 1.1$\footnotemark[1] & \cite{Hoeft:1972} \\
\noalign{\smallskip}\hline
\end{tabular}
\end{table}

\begin{table}
\caption{Comparison between spin-rotation coupling constants
calculated as described in the text and experimentally measured. The
label 1 refers to the less electronegative atom (K, Rb or Cs) and
the label 2 to the more electronegative one. The units are kHz.}
\label{tb:rc}       
\begin{tabular}{llllll}
\hline\noalign{\smallskip} Molecule & $c_{\rm 1}^{\rm Calc}$ &
 $c_{\rm 1}^{\rm Exp}$  & $c_{\rm 2}^{\rm Calc}$& $c_{\rm 2}^{\rm Exp}$ & {\rm Ref.} \\
\noalign{\smallskip}\hline\noalign{\smallskip}
$^{23}$Na$_{2}$  & 0.299 & 0.243 & 0.299 & 0.243 & \cite{Esbroeck:1985}\\
$^{39}$K$^{19}$F  & 0.235 & 0.270 & 17.5 & 10.7 & \cite{Bonczyk:1967}\\
$^{85}$Rb$^{19}$F & 0.598 & 0.498 & 16.1 & 10.6 & \cite{Cederberg:2006}\\
$^{85}$Rb$^{35}$Cl & 0.457 & 0.395 & 0.569 & 0.394 & \cite{Cederberg:2006a}\\
$^{133}$Cs$^{19}$F & 1.05 & 0.662 & 21.9 & 15.1 & \cite{Cederberg:1999}\\
\noalign{\smallskip}\hline
\end{tabular}
\end{table}

\begin{table}
\caption{Comparison between spin-spin coupling constants calculated
as described in the text and experimentally measured. The units are
kHz.}
\label{tb:sscc}       
\begin{tabular}{llllll}
\hline\noalign{\smallskip} Molecule & $c_{\rm 3}^{\rm Calc}$ &
 $c_{\rm 3}^{\rm Exp}$  & $c_{\rm 4}^{\rm Calc}$& $c_{\rm 4}^{\rm Exp}$ & {\rm Ref.}\\
\noalign{\smallskip}\hline\noalign{\smallskip}
$^{23}$Na$_{2}$  & 0.298 & 0.303 & 1.358 & 1.067 & \cite{Esbroeck:1985}\\
$^{39}$K$^{19}$F  & 0.470 & 0.540 & 0.032 & 0.030 & \cite{Bonczyk:1967}\\
$^{85}$Rb$^{19}$F & 0.751 & 0.797 & 0.151 & 0.237 & \cite{Cederberg:2006}\\
$^{85}$Rb$^{35}$Cl & 0.032 & 0.033 & 0.010 & 0.026 & \cite{Cederberg:2006a}\\
$^{133}$Cs$^{19}$F & 0.875 & 0.927 & 0.471 & 0.627 & \cite{Cederberg:1999}\\
\noalign{\smallskip}\hline
\end{tabular}
\end{table}

\begin{table}
\caption{Comparison between rotational g-factors calculated as
described in the text and experimentally measured.}
\label{tb:rgf}       
\begin{tabular}{llll}
\hline\noalign{\smallskip} Molecule & $g_{\rm r}^{\rm Calc}$ &
 $g_{\rm r}^{\rm Exp}$ & {\rm Ref.} \\
\noalign{\smallskip}\hline\noalign{\smallskip}
$^{23}$Na$_{2}$ & 0.0324 &0.0386 & \cite{Brooks:1963} \\
$^{39}$K$_{2}$ & 0.0247 & 0.0212 & \cite{Brooks:1963} \\
$^{23}$Na$^{39}$K  & 0.0253 & 0.0253 & \cite{Brooks:1972} \\
$^{85}$Rb$_{2}$ & 0.0082 & 0.0095  & \cite{Brooks:1963} \\
$^{133}$Cs$_{2}$ & 0.0051 & 0.0054 & \cite{Brooks:1963} \\
\noalign{\smallskip}\hline
\end{tabular}
\end{table}

Evaluation of hyperfine coupling constants requires a basis set
that properly describes the electron density near the nuclei.
Because of this, we employ all-electron basis sets rather than
valence basis sets with effective core potentials. However, for
core orbitals of heavy elements such as those considered here,
relativistic effects can be important. In the present work,
relativistic corrections were included by means of ZORA, the
two-component zero-order regular approximation
\cite{Lenthe:1993,Lenthe:1994,Lenthe:1999}, including
spin-orbit coupling as well as scalar effects (which are the
equivalent of Darwin and mass-velocity terms in the Breit-Pauli
Hamiltonian).

DFT generally performs well for calculations of electric
quadrupole coupling constants for main-group elements
\cite{Fedotov:1996,Bailey:1998,Bailey:1998a,Bailey:2000,Lenthe:2000,Hung:2003,Palmer:2007,Bischoff:2007,Behzadi:2007}.
Following most of these examples, we use the B3LYP functional
\cite{lee:1988,becke:1993} in our calculations with the QZ4P
basis set (a quadruple-$\zeta$ all-electron basis set with four
polarization functions).

Shielding tensors were evaluated using the KT2 functional
\cite{Keal:2003} with the same basis set and relativistic
correction as for the quadrupole coupling constants. For
calculation of shielding tensors of main-group atoms (H, C, N,
O and F), the performance of this functional is excellent, and
is better \cite{Keal:2004} than that of more popular
functionals such as BLYP \cite{becke:1988,lee:1988} and B3LYP.

Two nuclear magnetic moments can interact both directly
(through space) and indirectly (via the electron distribution).
The coupling constant for the direct interaction is
\cite{Bryce:2003,Vaara:2002}
\begin{equation}\label{eq:rdd}
    R_{\rm DD}=\frac{\mu_{0}}{4\pi} \frac{\mu_{\rm N}^2}{h} g_1g_2\langle
    R^{-3}\rangle,
\end{equation}
where $R$ is the internuclear distance. The indirect
interaction is represented by a tensor $\bm J$
\cite{Bryce:2003,Vaara:2002} with isotropic part $J_{\rm iso}$
and anisotropy $\Delta J = J_{\|} - J_{\bot}$. The coupling
constants $c_{3}$ and $c_{4}$ are related to the direct and
indirect components by \cite{Bryce:2003,Vaara:2002}
\begin{equation}\label{eq:ssc3}
c_{3}=R_{\rm DD}-\frac{\Delta J}{3}.
\end{equation}
and
\begin{equation}\label{eq:ssc4}
c_{4}=J_{\rm iso}
\end{equation}
In the present work, $c_{3}$ and $c_{4}$ were evaluated from
equations \ref{eq:rdd} to \ref{eq:ssc4} with $\langle R^{-3}
\rangle \simeq R_{\rm e}^{-3}$, where $R_{\rm e}$ is the
equilibrium distance. The components of $\bm J$ were calculated
using the same methods as for the quadrupole coupling
constants, except that the PBE \cite{perdew:1996} functional
was used. This functional produced results slightly closer to
the experimental measurements than KT2 for the molecules
considered in table \ref{tb:sscc} (although the differences
were small). BLYP performed well for all except Na$_2$, for
which it gave the wrong sign and order of magnitude; it also
gave qualitatively different results from PBE and KT2 for KRb
and RbCs.

ADF does not calculate spin-rotation constants directly.
However, the spin-rotation constants are given approximately by
\cite{Flygare:1964,Gierke:1972,Wasylishen:2000}
\begin{equation}\label{eq:src}
c_i\approx\frac{2m_{\rm e}B_v g_{i}}{m_{\rm p}}(\sigma_{i\|} - \sigma_{i\bot}) \quad\hbox{for\ } i=1,2,
\end{equation}
where $m_{\rm p}$ and $m_{\rm e}$ are the proton and electron
masses, $B_v$ is the rotational constant, $g_{i}$ is the
nuclear g-factor and $\sigma_{i\|} - \sigma_{i\bot}$ is the
anisotropy of the nuclear shielding tensor ${\bm \sigma}_i$.
Two approximations underlie this expression. First, a
quadrupole term has been neglected. Secondly, it was obtained
in the frame of the non-relativistic theory developed by
Flygare \cite{Flygare:1964}. However, previous studies
\cite{Cooke:2004} and our own results (see table \ref{tb:rc})
suggest that it can be applied reliably in the relativistic
case.

\begin{table*}
\caption{Nuclear properties and coupling constants for the
different isotopic species of the KRb molecule.}
\label{tb:KRb}       
\begin{tabular}{lllllll}
\hline\noalign{\smallskip} & $^{39}\rm{K}^{85}\rm{Rb}$ &
$^{39}\rm{K}^{87}\rm{Rb}$ & $^{40}\rm{K}^{85}\rm{Rb}$ &
 $^{40}\rm{K}^{87}\rm{Rb}$ & $^{41}\rm{K}^{85}\rm{Rb}$ & $^{41}\rm{K}^{87}\rm{Rb}$  \\
\noalign{\smallskip}\hline\noalign{\smallskip}
$I_{\rm K}$ & 3/2 & 3/2 & 4 & 4 & 3/2 & 3/2\\
$I_{\rm Rb}$ & 5/2 & 3/2 & 5/2 & 3/2 & 5/2 & 3/2\\
$g_{\rm K}$ & 0.261 & 0.261 & $-$0.324 & $-$0.324 & 0.143 & 0.143\\
$g_{\rm Rb}$ & 0.541 & 1.834 & 0.541 & 1.834 & 0.541 & 1.834\\
$B_v/{\rm GHz}$ & 1.142 & 1.134 & 1.123 & 1.114 & 1.104 & 1.096 \\
$(eQq)_{{\rm K}}/{\rm MHz}$ & $-$0.245 & $-$0.245 & 0.306 & 0.306 & $-$0.298 & $-$0.298 \\
$(eQq)_{{\rm Rb}}/{\rm MHz}$ & $-$3.142 & $-$1.520 & $-$3.142 & $-$1.520 & $-$3.142 & $-$1.520 \\
$\sigma_{{\rm K}}$(ppm) & 1321 & 1321 & 1321 & 1321 & 1321 & 1321 \\
$\sigma_{{\rm Rb}}$(ppm) & 3469 & 3469 & 3469 & 3469 & 3469 & 3469 \\
$c_{\rm K}/{\rm Hz}$ & 19.9 & 19.8 & $-$24.2 &$-$24.1 & 10.5 & 10.4 \\
$c_{\rm Rb}/{\rm Hz}$ & 127.0 & 427.5 & 124.8 & 420.1 & 122.8 & 413.1 \\
$c_{3}/{\rm Hz}$ & 11.5 & 38.9 & $-$14.2 & $-$48.2 & 6.3 & 21.3 \\
$c_{4}/{\rm Hz}$ & 482.5 & 1635.7 & $-$599.0 & $-$2030.4 & 264.3 & 896.2 \\
$g_{{\rm r}}$ & 0.0144 & 0.0142 & 0.0141 & 0.0140 & 0.0139 & 0.0138 \\
$\mu/{\rm D}$ & 0.76 & 0.76 & 0.76 & 0.76 & 0.76 & 0.76 \\
\noalign{\smallskip}\hline
\end{tabular}
\end{table*}

Lastly, the rotational g-factors were evaluated with the DALTON
program using the KT2 functional and the all-electron basis
sets of Huzinaga and coworkers
\cite{Huzinaga:1990,Huzinaga:1993}. Again, the choice of the
functional is based on its reliability for this molecular
property \cite{Wilson:2005}. No relativistic corrections were
included in this case. Previous calculations
\cite{Enevoldsen:2001} for hydrogen halides and noble gas
hydride cations including atoms as heavy as I and Xe suggest
that relativistic corrections are relatively small for
rotational g-factors (less than $5\%$ of the non-relativistic
value).

The coupling constants obtained for KRb and RbCs are given in
tables \ref{tb:KRb} and \ref{tb:RbCs}. All the calculations
were carried out at the equilibrium geometries, $R_{\rm
e}=4.07$~\AA\ for KRb \cite{Ross:1990} and $R_{\rm
e}=4.37$~\AA\ for RbCs \cite{Kato:1983}. This neglects small
corrections due to vibrational averaging even for $v=0$, but
nevertheless gives results that are qualitatively valid for any
low-lying vibrational state. ADF generally gives coupling
constants for only one isotopic species, but the others may be
obtained by simple scaling. The nuclear quadrupole coupling
constants scale with the nuclear quadrupoles $Q_i$, the
spin-spin coupling constants with the product of nuclear
g-factors $g_i g_j$, and the spin-rotation coupling constant
with the product of $g_i$ and the rotational constant $B_v$.
The rotational g-factor scales in a more complicated way that
depends on $B_v$ and the shift of the center of mass
\cite{Lawrence:1963}.

\begin{table}
\caption{Nuclear properties and coupling constants for the
different isotopic species of the RbCs molecule.}
\label{tb:RbCs}       
\begin{tabular}{lll}
\hline\noalign{\smallskip}
& $^{85}\rm{Rb}^{133}\rm{Cs}$ & $^{87}\rm{Rb}^{133}\rm{Cs}$  \\
\noalign{\smallskip}\hline\noalign{\smallskip}
$I_{\rm Rb}$ & 5/2 & 3/2 \\
$I_{\rm Cs}$ & 7/2 & 7/2 \\
$g_{\rm Rb}$ & 0.541 & 1.834 \\
$g_{\rm Cs}$ & 0.738 & 0.738 \\
$B_v/{\rm GHz}$ & 0.511 & 0.504 \\
$(eQq)_{{\rm Rb}}/{\rm MHz}$ & $-$1.803 & $-$0.872 \\
$(eQq)_{{\rm Cs}}/{\rm MHz}$ & 0.051 & 0.051 \\
$\sigma_{{\rm Rb}}$(ppm) & 3531 & 3531 \\
$\sigma_{{\rm Cs}}$(ppm) & 6367 & 6367 \\
$c_{\rm Rb}/{\rm Hz}$ & 29.4 & 98.4 \\
$c_{\rm Cs}/{\rm Hz}$ & 196.8 & 194.1 \\
$c_{3}/{\rm Hz}$ & 56.8 & 192.4 \\
$c_{4}/{\rm Hz}$ & 5116.6 & 17345.4 \\
$g_{\rm r}$ & 0.0063 & 0.0062 \\
$\mu/{\rm D}$ & 1.25 & 1.25 \\
\noalign{\smallskip}\hline
\end{tabular}
\end{table}

\section{Hyperfine energy levels} \label{sec:Molenlev}

We calculate the hyperfine levels by diagonalizing the complete
Hamiltonian of equations \ref{eq:Hrot} to \ref{eq:Hz} in a
basis set of angular momentum functions. We employ three
different basis sets,
\begin{eqnarray}
  \label{eq:ub} |I_{1} M_{1} I_{2} M_{2} N M_{N}\rangle & & (\mbox{uncoupled basis}); \\
  \label{eq:cb} |(I_{1} I_{2}) I M_{I} N M_{N}\rangle & & (\mbox{spin-coupled basis}); \\
  \label{eq:tcb} |(I_{1} I_{2}) I  N F M_{F}\rangle & & (\mbox{fully coupled basis}).
\end{eqnarray}
Here $I$ and $F$ are quantum numbers for the total nuclear spin
and total angular momentum and $M_{I}$ and $M_{F}$ represent
their projections onto the $Z$ axis defined by the external
field. We consider here only cases in which only one field,
electric or magnetic, in present. The matrix elements
corresponding to the different terms of the Hamiltonian in each
of the basis sets are calculated through standard angular
momentum techniques \cite{Zare}.

The use of three basis sets rather than one helps in assigning
quantum numbers to the energy levels. Although the Hamiltonian
matrix is not diagonal in any of the basis sets employed, it is
usually closer to diagonal for one basis than for the others.
When one coefficient of an eigenvector is much larger than the
others, it is possible to assign approximate quantum numbers to
the state concerned. However, different basis sets achieve this
in different field regimes.

\subsection{Zeeman splitting for rotational ground-state molecules ($N=0$)}
\label{sec:zeemann0}

\begin{figure}
  \resizebox{1.0\hsize}{!}{\includegraphics*{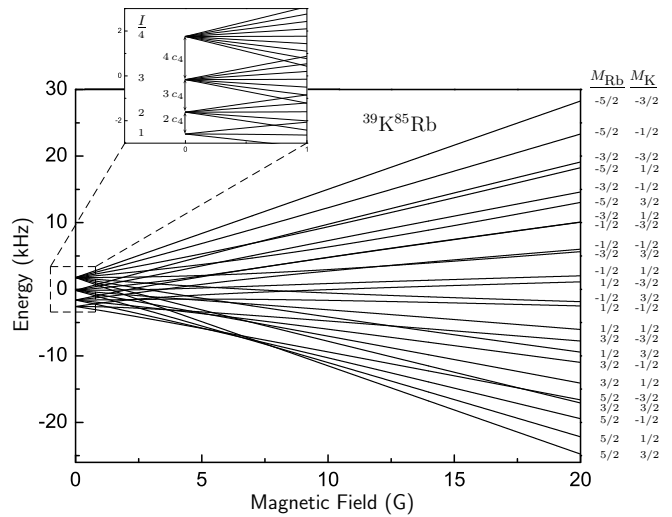}}
  \caption{\label{fig:01}%
    Zeeman levels for $^{39}{\rm K}^{85}{\rm Rb}(v=0,N=0)$.}
\end{figure}

\begin{figure}
  \resizebox{1.0\hsize}{!}{\includegraphics*{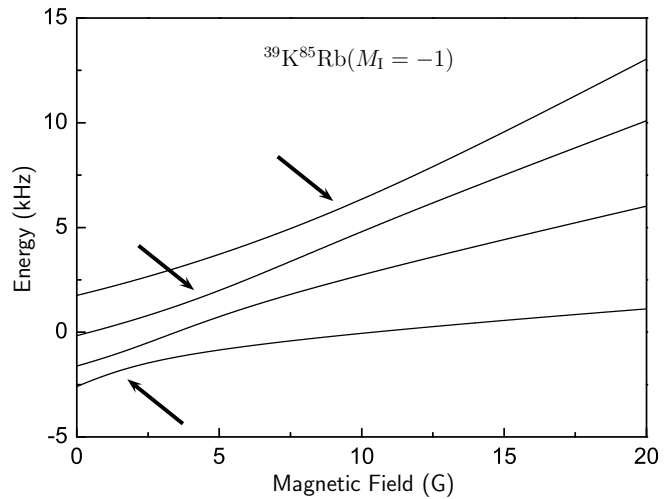}}
  \caption{\label{fig:02}%
     Zeeman splitting and avoided crossings (indicated with arrows)
     for the $M_{I}=-1$ levels of
     $^{39}{\rm K}^{85}{\rm Rb}(v=0,N=0)$.}
\end{figure}

Figure \ref{fig:01} shows the Zeeman splittings for energy
levels of $^{39}{\rm K}^{85}{\rm Rb}$ with $N=0$. The
splittings are dominated by the scalar nuclear spin-spin
interaction and the nuclear Zeeman effect, which are the only
terms in the Hamiltonian with matrix elements diagonal in $N$
for $N=0$. It should be noted that the scalar spin-spin
coupling is entirely mediated by the electron distribution, and
has no contribution from the direct dipolar interaction. In the
absence of external fields, the energy levels are split into
groups labeled by the total nuclear spin $I$. For small
magnetic fields $B$, $I$ remains a nearly good quantum number
and the levels split according to the value of its projection
$M_{I}$ (which in this case coincides with the projection of
the total angular momentum, which is always a good quantum
number). Energy levels corresponding to the same value of
$M_{I}$ display avoided crossings as a function of the field as
shown in figure \ref{fig:02}. For fields well above the
crossings (which are at 2 to 10 G in this case), $I$ is
destroyed and the good quantum numbers are $M_{\rm Rb}$ and
$M_{\rm K}$. Since both nuclear g-factors are positive for
$^{39}{\rm K}^{85}{\rm Rb}$, states where both projections are
positive are high-field-seeking and those where both are
negative are low-field-seeking.

%
%

%
%

Although the splittings at low fields are dominated by the
scalar spin-spin coupling, there are several terms in the
Hamiltonian that are off-diagonal in $N$. The energies are
therefore obtained by diagonalizing a full matrix that includes
enough rotational levels for convergence. For the Zeeman
effect, the only off-diagonal terms involving $N=0$ are the
electric quadrupole coupling and the tensor spin-spin coupling,
both of which are small. Convergence for $N=0$ is achieved with
$N_{\rm max}=2$ and the splittings obtained differ from those
calculated with only $N=0$ by less than 1\%. For the Stark
effect, however, the Stark term itself mixes $N=0$ states with
$N>0$. Terms off-diagonal in $N$ are then very important and
much larger basis sets are needed.

The scalar spin-spin interaction for $N=0$ is diagonal in the
spin-coupled and fully coupled basis sets,
\begin{eqnarray}
\nonumber \lefteqn { \langle N=0 (I_{1} I_{2}) I M_{I}|
c_{4}\,\bm{I}_{1}\cdot\bm{I}_{2}| N=0 (I_{1} I_{2}) I M_{I}
   \rangle = }  \\
\nonumber \lefteqn { \langle N=0 (I_{1} I_{2}) I F M_{F}|
c_{4}\,\bm{I}_{1}\cdot\bm{I}_{2}| N=0 (I_{1} I_{2}) I F M_{F}
   \rangle = }  \\
&& \label{eq:Hescj0} \frac{1}{2} c_{4}
[I(I+1)-I_{1}(I_{1}+1)-I_{2}(I_{2}+1)].
\end{eqnarray}
The nuclear Zeeman Hamiltonian is diagonal in the uncoupled basis
set, with nonzero elements given by
\begin{equation}\label{eq:HZej0}
-[g_{\rm 1}M_{1}(1-\sigma_1)+g_{\rm 2}M_{2}(1-\sigma_2)]\mu_{\rm N}B.
\end{equation}
The splitting pattern is therefore determined by the allowed
values of the total nuclear spin quantum number $I$ and by the
magnitudes and signs of the scalar spin-spin coupling constant
$c_{4}$ and the rotational g-factors. The nuclear shielding
constants $\sigma_i$ are only a few parts per thousand. For
large values of the magnetic field, where the nuclear Zeeman
effect is the dominant term in the Hamiltonian, the magnetic
moment (gradient of the energy with respect to $B$) is close to
$-(g_{\rm 1}M_{1}+g_{\rm 2}M_{2})\mu_{\rm N}$.

%
%
\begin{figure}
  \resizebox{1.0\hsize}{!}{\includegraphics*{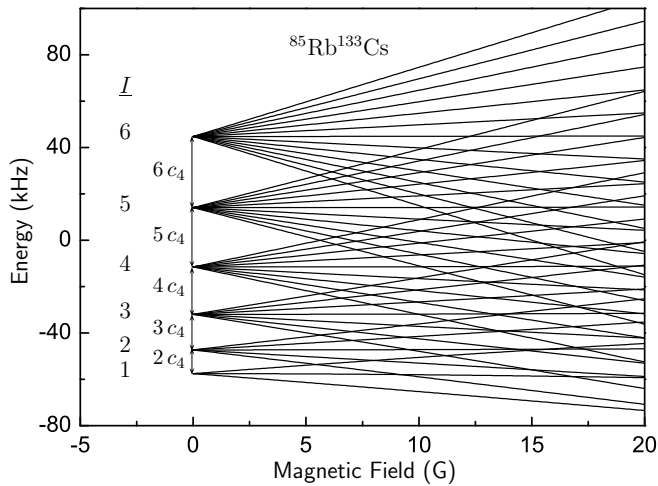}}
  \caption{\label{fig:03}%
    Zeeman levels for $^{85}{\rm Rb}^{133}{\rm Cs}(v=0,N=0)$.}
\end{figure}

The Zeeman splittings for $^{85}$Rb$^{133}$ Cs are shown in
figure \ref{fig:03}. They are qualitatively similar to those
for $^{39}$K$^{85}$Rb, except that the range of $I$ is
different and the spin-spin coupling constant $c_4$ is
significantly larger. Because of this, $I$ remains a good
quantum number up to significantly higher magnetic fields. At
high fields, once the magnitude of the scalar spin-spin
interaction can be neglected compared to the Zeeman effect,
$M_{\rm Rb}$ and $M_{\rm Cs}$ become good quantum numbers.

The splitting patterns for other KRb and RbCs isotopologs are
qualitatively similar to those discussed above and the
corresponding figures are available as supplementary online
material. The spin-spin coupling constant and the potassium
g-factor are negative for $^{40}$K$^{85}$Rb and
$^{40}$K$^{87}$Rb. The sign of $c_{4}$ determines whether the
lowest zero-field energy corresponds to the highest or lowest
value of $I$. In general the fields where the avoided crossings
occur and above which $M_1$ and $M_2$ become good quantum
numbers scale with $|c_4/(g_1-g_2)|$. When $g_1$ and $g_2$ are
equal, as in homonuclear dimers, there are no avoided crossings
for $N=0$ and the $I$ quantum number is conserved even at high
fields.

\subsection{Stark splitting for rotational ground-state molecules ($N=0$)}
\label{sec:starkn0}

\begin{figure}
  \resizebox{1.0\hsize}{!}{\includegraphics*{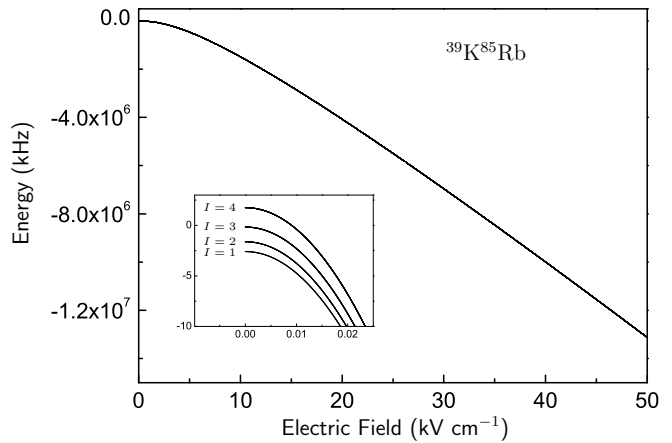}}
  \caption{\label{fig:04}%
    Stark effect on energy levels of $^{39}{\rm K}^{85}{\rm Rb}$
    correlating with $(v=0,N=0)$ for electric fields up to 50 kV/cm.}
\end{figure}

\begin{figure}
  \resizebox{1.0\hsize}{!}{\includegraphics*{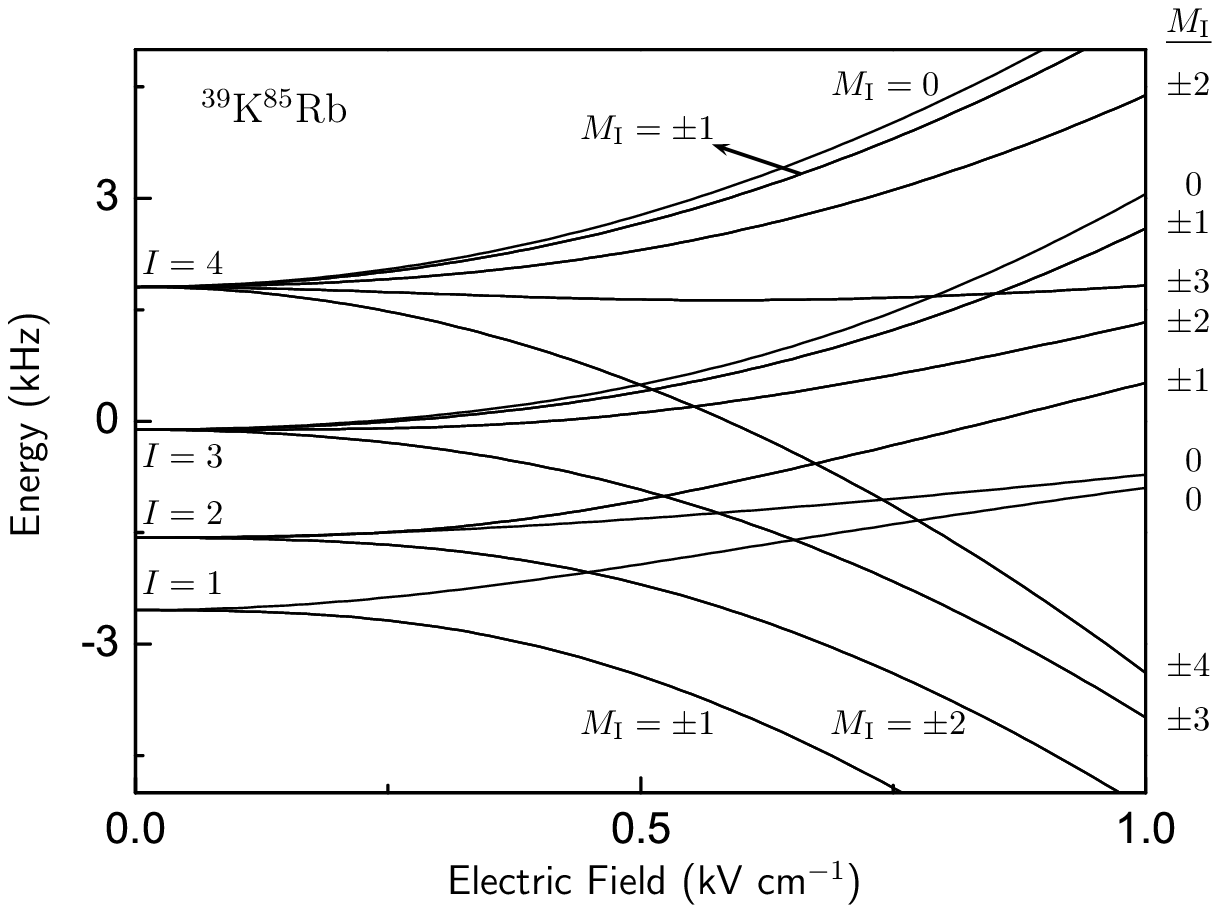}}
  \caption{\label{fig:05}%
    Stark splitting for energy levels of $^{39}{\rm K}^{85}{\rm Rb}$
    correlating with $(v=0,N=0)$ for electric fields up to 1 kV/cm.
    The levels are shown relative to their field-dependent average
    energy.}
\end{figure}

\begin{figure}
  \resizebox{1.0\hsize}{!}{\includegraphics*{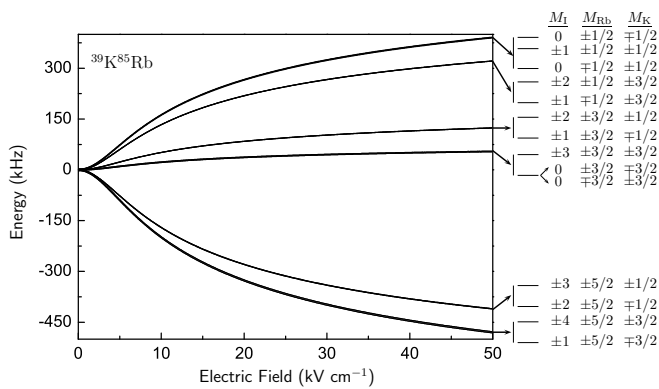}}
  \caption{\label{fig:06}%
    Stark splitting for energy levels of $^{39}{\rm K}^{85}{\rm Rb}$
    correlating with $(v=0,N=0)$ for electric fields up to 50 kV/cm.
    The levels are shown relative to their field-dependent average
    energy.}
\end{figure}

The Stark effect for levels of $^{39}{\rm K}^{85}{\rm Rb}$
correlating with $N=0$ is shown in figure \ref{fig:04} to
\ref{fig:06}. Corresponding figures for the remaining
isotopologs of KRb and RbCs are available as additional online
material. The Stark effect is quadratic at low fields but
becomes linear at high fields, as is usual for diatomic
molecules in $\Sigma$ states \cite{Townes}. This arises from
mixing between different rotational levels: while in the Zeeman
case this mixing is very weak and is exclusively due to
hyperfine terms, in the Stark case it is strong and is caused
directly by the electric field. At low fields the mixing is
weak and can be treated by second-order perturbation theory,
giving rise to a quadratic Stark effect. However, as the field
increases the mixing becomes increasingly important: the $N=1$
basis functions contribute around 25\% at 10 kV/cm and 40\% at
20 kV/cm. Eventually the molecule becomes fully oriented by the
field and the linear Stark effect overcomes the quadratic
effect. The mixing also has numerical consequences as the
number of rotational levels required for convergence increases
with field: for example, calculations at 50 kV/cm require
$N_{\rm max}=6$.

%
%

%
%

%
%

The magnitude of the Stark shift in figure \ref{fig:04}
obscures the splittings between hyperfine levels. Figure
\ref{fig:05} therefore shows the levels correlating with $N=0$
relative to their average energy, for fields up to 1 kV/cm. As
expected, each zero-field level splits into $I+1$ components
labeled by the different possible values of $|M_I|$. For
$|M_I|>0$ the levels exist in degenerate pairs corresponding to
changing the sign of $M_1$ {\em and} $M_2$. However, changing
the sign of {\em one} of $M_1$ and $M_2$ produces a different
state with a different value of $|M_I|$. For $M_I=0$ there is
an extra symmetry corresponding to reflection in a plane
containing the electric field vector.

At higher field, as shown in figure \ref{fig:06}, the
projections of the individual nuclear spins become well-defined
as well as their sum. At sufficiently large fields the
splittings approach a limiting value as the molecules become
strongly oriented along the field direction. In this limit the
splittings are mostly determined by the nuclear quadrupole
coupling constants, with relatively small contributions from
the magnetic hyperfine terms.

\section{Conclusion}
\label{sec:conc}

We have investigated the hyperfine level splittings expected
for alkali metal dimers in their rotational ground state in the
presence of electric and magnetic fields. We have carried out
density-functional calculations of the electronic structure of
RbK and RbCs at the equilibrium geometry of the ground
$^1\Sigma$ state and evaluated all the hyperfine coupling
constants necessary to calculate energy level patterns. For
nonrotating states, the zero-field splittings between hyperfine
states range from a few kHz for isotopologs of KRb to a few
tens of kHz for isotopologs of RbCs. They are dominated by the
electron-mediated contribution to the nuclear spin-spin
coupling. The results will be valuable in designing laser-based
schemes to produce ultracold molecules in their absolute ground
states in applied fields.

\section*{Acknowledgments}
The authors are grateful to EPSRC for funding of the
collaborative project QuDipMol under the ESF EUROCORES
Programme EuroQUAM and to the UK National Centre for
Computational Chemistry Software for computer facilities.

\bibliography{jesus,../../all}

\begin{thebibliography}{72}
\expandafter\ifx\csname natexlab\endcsname\relax\def\natexlab#1{#1}\fi
\expandafter\ifx\csname bibnamefont\endcsname\relax
  \def\bibnamefont#1{#1}\fi
\expandafter\ifx\csname bibfnamefont\endcsname\relax
  \def\bibfnamefont#1{#1}\fi
\expandafter\ifx\csname citenamefont\endcsname\relax
  \def\citenamefont#1{#1}\fi
\expandafter\ifx\csname url\endcsname\relax
  \def\url#1{\texttt{#1}}\fi
\expandafter\ifx\csname urlprefix\endcsname\relax\def\urlprefix{URL }\fi
\providecommand{\bibinfo}[2]{#2}
\providecommand{\eprint}[2][]{\url{#2}}

\bibitem[{\citenamefont{Hutson and Sold\'{a}n}(2006)}]{Hutson:IRPC:2006}
\bibinfo{author}{\bibfnamefont{J.~M.} \bibnamefont{Hutson}} \bibnamefont{and}
  \bibinfo{author}{\bibfnamefont{P.}~\bibnamefont{Sold\'{a}n}},
  \bibinfo{journal}{Int. Rev. Phys. Chem.} \textbf{\bibinfo{volume}{25}},
  \bibinfo{pages}{497} (\bibinfo{year}{2006}).

\bibitem[{\citenamefont{Jones et~al.}(2006)\citenamefont{Jones, Tiesinga, Lett,
  and Julienne}}]{Jones:RMP:2006}
\bibinfo{author}{\bibfnamefont{K.~M.} \bibnamefont{Jones}},
  \bibinfo{author}{\bibfnamefont{E.}~\bibnamefont{Tiesinga}},
  \bibinfo{author}{\bibfnamefont{P.~D.} \bibnamefont{Lett}}, \bibnamefont{and}
  \bibinfo{author}{\bibfnamefont{P.~S.} \bibnamefont{Julienne}},
  \bibinfo{journal}{Rev. Mod. Phys.} \textbf{\bibinfo{volume}{78}},
  \bibinfo{pages}{483} (\bibinfo{year}{2006}).

\bibitem[{\citenamefont{K\"{o}hler et~al.}(2006)\citenamefont{K\"{o}hler,
  Goral, and Julienne}}]{Koehler:RMP:2006}
\bibinfo{author}{\bibfnamefont{T.}~\bibnamefont{K\"{o}hler}},
  \bibinfo{author}{\bibfnamefont{K.}~\bibnamefont{Goral}}, \bibnamefont{and}
  \bibinfo{author}{\bibfnamefont{P.~S.} \bibnamefont{Julienne}},
  \bibinfo{journal}{Rev. Mod. Phys.} \textbf{\bibinfo{volume}{78}},
  \bibinfo{pages}{1311} (\bibinfo{year}{2006}).

\bibitem[{\citenamefont{Baranov et~al.}(2002)\citenamefont{Baranov, Dobrek,
  G\'{o}ral, Santos, and Lewenstein}}]{Baranov:2002}
\bibinfo{author}{\bibfnamefont{M.}~\bibnamefont{Baranov}},
  \bibinfo{author}{\bibfnamefont{{\L}.}~\bibnamefont{Dobrek}},
  \bibinfo{author}{\bibfnamefont{K.}~\bibnamefont{G\'{o}ral}},
  \bibinfo{author}{\bibfnamefont{L.}~\bibnamefont{Santos}}, \bibnamefont{and}
  \bibinfo{author}{\bibfnamefont{M.}~\bibnamefont{Lewenstein}},
  \bibinfo{journal}{Phys. Scr.} \textbf{\bibinfo{volume}{T102}},
  \bibinfo{pages}{74} (\bibinfo{year}{2002}).

\bibitem[{\citenamefont{DeMille}(2002)}]{DeMille:2002}
\bibinfo{author}{\bibfnamefont{D.}~\bibnamefont{DeMille}},
  \bibinfo{journal}{Phys. Rev. Lett.} \textbf{\bibinfo{volume}{88}},
  \bibinfo{pages}{067901} (\bibinfo{year}{2002}).

\bibitem[{\citenamefont{Jochim et~al.}(2003)\citenamefont{Jochim, Bartenstein,
  Altmeyer, Hendl, Riedl, Chin, Denschlag, and Grimm}}]{Jochim:Li2BEC:2003}
\bibinfo{author}{\bibfnamefont{S.}~\bibnamefont{Jochim}},
  \bibinfo{author}{\bibfnamefont{M.}~\bibnamefont{Bartenstein}},
  \bibinfo{author}{\bibfnamefont{A.}~\bibnamefont{Altmeyer}},
  \bibinfo{author}{\bibfnamefont{G.}~\bibnamefont{Hendl}},
  \bibinfo{author}{\bibfnamefont{S.}~\bibnamefont{Riedl}},
  \bibinfo{author}{\bibfnamefont{C.}~\bibnamefont{Chin}},
  \bibinfo{author}{\bibfnamefont{J.~H.} \bibnamefont{Denschlag}},
  \bibnamefont{and} \bibinfo{author}{\bibfnamefont{R.}~\bibnamefont{Grimm}},
  \bibinfo{journal}{Science} \textbf{\bibinfo{volume}{302}},
  \bibinfo{pages}{2101} (\bibinfo{year}{2003}).

\bibitem[{\citenamefont{Zwierlein et~al.}(2003)\citenamefont{Zwierlein, Stan,
  Schunck, Raupach, Gupta, Hadzibabic, and Ketterle}}]{Zwierlein:2003}
\bibinfo{author}{\bibfnamefont{M.~W.} \bibnamefont{Zwierlein}},
  \bibinfo{author}{\bibfnamefont{C.~A.} \bibnamefont{Stan}},
  \bibinfo{author}{\bibfnamefont{C.~H.} \bibnamefont{Schunck}},
  \bibinfo{author}{\bibfnamefont{S.~M.~F.} \bibnamefont{Raupach}},
  \bibinfo{author}{\bibfnamefont{S.}~\bibnamefont{Gupta}},
  \bibinfo{author}{\bibfnamefont{Z.}~\bibnamefont{Hadzibabic}},
  \bibnamefont{and} \bibinfo{author}{\bibfnamefont{W.}~\bibnamefont{Ketterle}},
  \bibinfo{journal}{Phys. Rev. Lett.} \textbf{\bibinfo{volume}{91}},
  \bibinfo{pages}{250401} (\bibinfo{year}{2003}).

\bibitem[{\citenamefont{Greiner et~al.}(2003)\citenamefont{Greiner, Regal, and
  Jin}}]{Greiner:2003}
\bibinfo{author}{\bibfnamefont{M.}~\bibnamefont{Greiner}},
  \bibinfo{author}{\bibfnamefont{C.~A.} \bibnamefont{Regal}}, \bibnamefont{and}
  \bibinfo{author}{\bibfnamefont{D.~S.} \bibnamefont{Jin}},
  \bibinfo{journal}{Nature} \textbf{\bibinfo{volume}{426}},
  \bibinfo{pages}{537} (\bibinfo{year}{2003}).

\bibitem[{\citenamefont{Petrov et~al.}(2004)\citenamefont{Petrov, Salomon, and
  Shlyapnikov}}]{Petrov:2004}
\bibinfo{author}{\bibfnamefont{D.~S.} \bibnamefont{Petrov}},
  \bibinfo{author}{\bibfnamefont{C.}~\bibnamefont{Salomon}}, \bibnamefont{and}
  \bibinfo{author}{\bibfnamefont{G.~V.} \bibnamefont{Shlyapnikov}},
  \bibinfo{journal}{Phys. Rev. Lett.} \textbf{\bibinfo{volume}{93}},
  \bibinfo{pages}{090404} (\bibinfo{year}{2004}).

\bibitem[{\citenamefont{Herbig et~al.}(2003)\citenamefont{Herbig, Kraemer,
  Mark, Weber, Chin, N\"{a}gerl, and Grimm}}]{Herbig:2003}
\bibinfo{author}{\bibfnamefont{J.}~\bibnamefont{Herbig}},
  \bibinfo{author}{\bibfnamefont{T.}~\bibnamefont{Kraemer}},
  \bibinfo{author}{\bibfnamefont{M.}~\bibnamefont{Mark}},
  \bibinfo{author}{\bibfnamefont{T.}~\bibnamefont{Weber}},
  \bibinfo{author}{\bibfnamefont{C.}~\bibnamefont{Chin}},
  \bibinfo{author}{\bibfnamefont{H.~C.} \bibnamefont{N\"{a}gerl}},
  \bibnamefont{and} \bibinfo{author}{\bibfnamefont{R.}~\bibnamefont{Grimm}},
  \bibinfo{journal}{Science} \textbf{\bibinfo{volume}{301}},
  \bibinfo{pages}{1510} (\bibinfo{year}{2003}).

\bibitem[{\citenamefont{Sold\'{a}n et~al.}(2002)\citenamefont{Sold\'{a}n,
  Cvita\v{s}, Hutson, Honvault, and Launay}}]{Soldan:2002}
\bibinfo{author}{\bibfnamefont{P.}~\bibnamefont{Sold\'{a}n}},
  \bibinfo{author}{\bibfnamefont{M.~T.} \bibnamefont{Cvita\v{s}}},
  \bibinfo{author}{\bibfnamefont{J.~M.} \bibnamefont{Hutson}},
  \bibinfo{author}{\bibfnamefont{P.}~\bibnamefont{Honvault}}, \bibnamefont{and}
  \bibinfo{author}{\bibfnamefont{J.~M.} \bibnamefont{Launay}},
  \bibinfo{journal}{Phys. Rev. Lett.} \textbf{\bibinfo{volume}{89}},
  \bibinfo{pages}{153201} (\bibinfo{year}{2002}).

\bibitem[{\citenamefont{Hutson and Sold\'{a}n}(2007)}]{Hutson:IRPC:2007}
\bibinfo{author}{\bibfnamefont{J.~M.} \bibnamefont{Hutson}} \bibnamefont{and}
  \bibinfo{author}{\bibfnamefont{P.}~\bibnamefont{Sold\'{a}n}},
  \bibinfo{journal}{Int. Rev. Phys. Chem.} \textbf{\bibinfo{volume}{26}},
  \bibinfo{pages}{1} (\bibinfo{year}{2007}).

\bibitem[{\citenamefont{Winkler et~al.}(2007)\citenamefont{Winkler, Lang,
  Thalhammer, van~der Straten, Grimm, Denschlag, Daley, Kantian, B\"{u}chler,
  and Zoller}}]{Winkler:2007}
\bibinfo{author}{\bibfnamefont{K.}~\bibnamefont{Winkler}},
  \bibinfo{author}{\bibfnamefont{F.}~\bibnamefont{Lang}},
  \bibinfo{author}{\bibfnamefont{G.}~\bibnamefont{Thalhammer}},
  \bibinfo{author}{\bibfnamefont{P.}~\bibnamefont{van~der Straten}},
  \bibinfo{author}{\bibfnamefont{R.}~\bibnamefont{Grimm}},
  \bibinfo{author}{\bibfnamefont{J.~H.} \bibnamefont{Denschlag}},
  \bibinfo{author}{\bibfnamefont{A.~J.} \bibnamefont{Daley}},
  \bibinfo{author}{\bibfnamefont{A.}~\bibnamefont{Kantian}},
  \bibinfo{author}{\bibfnamefont{H.~P.} \bibnamefont{B\"{u}chler}},
  \bibnamefont{and} \bibinfo{author}{\bibfnamefont{P.}~\bibnamefont{Zoller}},
  \bibinfo{journal}{Phys. Rev. Lett.} \textbf{\bibinfo{volume}{98}},
  \bibinfo{pages}{043201} (\bibinfo{year}{2007}).

\bibitem[{\citenamefont{Ospelkaus et~al.}(2008)\citenamefont{Ospelkaus, Pe'er,
  Ni, Zirbel, Neyenhuis, Kotochigova, Julienne, Ye, and Jin}}]{Ospelkaus:2008}
\bibinfo{author}{\bibfnamefont{S.}~\bibnamefont{Ospelkaus}},
  \bibinfo{author}{\bibfnamefont{A.}~\bibnamefont{Pe'er}},
  \bibinfo{author}{\bibfnamefont{K.-K.} \bibnamefont{Ni}},
  \bibinfo{author}{\bibfnamefont{J.~J.} \bibnamefont{Zirbel}},
  \bibinfo{author}{\bibfnamefont{B.}~\bibnamefont{Neyenhuis}},
  \bibinfo{author}{\bibfnamefont{S.}~\bibnamefont{Kotochigova}},
  \bibinfo{author}{\bibfnamefont{P.~S.} \bibnamefont{Julienne}},
  \bibinfo{author}{\bibfnamefont{J.}~\bibnamefont{Ye}}, \bibnamefont{and}
  \bibinfo{author}{\bibfnamefont{D.~S.} \bibnamefont{Jin}},
  \bibinfo{journal}{arXiv:physics/0802.1093}  (\bibinfo{year}{2008}).

\bibitem[{\citenamefont{Danzl et~al.}(2008)\citenamefont{Danzl, Haller,
  Gustavsson, Mark, Hart, Bouloufa, Dulieu, Ritsch, and N\"agerl}}]{Danzl:2008}
\bibinfo{author}{\bibfnamefont{J.~G.} \bibnamefont{Danzl}},
  \bibinfo{author}{\bibfnamefont{E.}~\bibnamefont{Haller}},
  \bibinfo{author}{\bibfnamefont{M.}~\bibnamefont{Gustavsson}},
  \bibinfo{author}{\bibfnamefont{M.~J.} \bibnamefont{Mark}},
  \bibinfo{author}{\bibfnamefont{R.}~\bibnamefont{Hart}},
  \bibinfo{author}{\bibfnamefont{N.}~\bibnamefont{Bouloufa}},
  \bibinfo{author}{\bibfnamefont{O.}~\bibnamefont{Dulieu}},
  \bibinfo{author}{\bibfnamefont{H.}~\bibnamefont{Ritsch}}, \bibnamefont{and}
  \bibinfo{author}{\bibfnamefont{H.-C.} \bibnamefont{N\"agerl}},
  \bibinfo{journal}{arXiv:physics/0806.2284}  (\bibinfo{year}{2008}).

\bibitem[{\citenamefont{Mark et~al.}(2007{\natexlab{a}})\citenamefont{Mark,
  Kraemer, Waldburger, Herbig, Chin, N\"agerl, and Grimm}}]{Mark:stuck:2007}
\bibinfo{author}{\bibfnamefont{M.}~\bibnamefont{Mark}},
  \bibinfo{author}{\bibfnamefont{T.}~\bibnamefont{Kraemer}},
  \bibinfo{author}{\bibfnamefont{P.}~\bibnamefont{Waldburger}},
  \bibinfo{author}{\bibfnamefont{J.}~\bibnamefont{Herbig}},
  \bibinfo{author}{\bibfnamefont{C.}~\bibnamefont{Chin}},
  \bibinfo{author}{\bibfnamefont{H.-C.} \bibnamefont{N\"agerl}},
  \bibnamefont{and} \bibinfo{author}{\bibfnamefont{R.}~\bibnamefont{Grimm}},
  \bibinfo{journal}{Phys. Rev. Lett.} \textbf{\bibinfo{volume}{99}},
  \bibinfo{pages}{113201} (\bibinfo{year}{2007}{\natexlab{a}}).

\bibitem[{\citenamefont{Mark et~al.}(2007{\natexlab{b}})\citenamefont{Mark,
  Ferlaino, Knoop, Kraemer, Chin, N\"agerl, and Grimm}}]{Mark:spect:2007}
\bibinfo{author}{\bibfnamefont{M.}~\bibnamefont{Mark}},
  \bibinfo{author}{\bibfnamefont{F.}~\bibnamefont{Ferlaino}},
  \bibinfo{author}{\bibfnamefont{S.}~\bibnamefont{Knoop}},
  \bibinfo{author}{\bibfnamefont{T.}~\bibnamefont{Kraemer}},
  \bibinfo{author}{\bibfnamefont{C.}~\bibnamefont{Chin}},
  \bibinfo{author}{\bibfnamefont{H.-C.} \bibnamefont{N\"agerl}},
  \bibnamefont{and} \bibinfo{author}{\bibfnamefont{R.}~\bibnamefont{Grimm}},
  \bibinfo{journal}{Phys. Rev. A} \textbf{\bibinfo{volume}{76}},
  \bibinfo{pages}{042514} (\bibinfo{year}{2007}{\natexlab{b}}).

\bibitem[{\citenamefont{Chin et~al.}(2004)\citenamefont{Chin, Vuleti\'c,
  Kerman, Chu, Tiesinga, Leo, and Williams}}]{Chin:cs2-fesh:2004}
\bibinfo{author}{\bibfnamefont{C.}~\bibnamefont{Chin}},
  \bibinfo{author}{\bibfnamefont{V.}~\bibnamefont{Vuleti\'c}},
  \bibinfo{author}{\bibfnamefont{A.~J.} \bibnamefont{Kerman}},
  \bibinfo{author}{\bibfnamefont{S.}~\bibnamefont{Chu}},
  \bibinfo{author}{\bibfnamefont{E.}~\bibnamefont{Tiesinga}},
  \bibinfo{author}{\bibfnamefont{P.~J.} \bibnamefont{Leo}}, \bibnamefont{and}
  \bibinfo{author}{\bibfnamefont{C.~J.} \bibnamefont{Williams}},
  \bibinfo{journal}{Phys. Rev. A} \textbf{\bibinfo{volume}{70}},
  \bibinfo{pages}{032701} (\bibinfo{year}{2004}).

\bibitem[{\citenamefont{Hutson et~al.}(2008)\citenamefont{Hutson, Tiesinga, and
  Julienne}}]{Hutson:Cs2:2008}
\bibinfo{author}{\bibfnamefont{J.~M.} \bibnamefont{Hutson}},
  \bibinfo{author}{\bibfnamefont{E.}~\bibnamefont{Tiesinga}}, \bibnamefont{and}
  \bibinfo{author}{\bibfnamefont{P.~S.} \bibnamefont{Julienne}},
  \bibinfo{journal}{arXiv:physics/0806.2583}  (\bibinfo{year}{2008}).

\bibitem[{\citenamefont{Ramsey}(1952)}]{Ramsey:1952}
\bibinfo{author}{\bibfnamefont{N.~F.} \bibnamefont{Ramsey}},
  \bibinfo{journal}{Phys. Rev.} \textbf{\bibinfo{volume}{85}},
  \bibinfo{pages}{60} (\bibinfo{year}{1952}).

\bibitem[{\citenamefont{Brown and Carrington}(2003)}]{Brown}
\bibinfo{author}{\bibfnamefont{J.~M.} \bibnamefont{Brown}} \bibnamefont{and}
  \bibinfo{author}{\bibfnamefont{A.}~\bibnamefont{Carrington}},
  \emph{\bibinfo{title}{Rotational Spectroscopy of Diatomic Molecules}}
  (\bibinfo{publisher}{Cambridge University Press, Cambridge},
  \bibinfo{year}{2003}).

\bibitem[{\citenamefont{Bryce and Wasylishen}(2003)}]{Bryce:2003}
\bibinfo{author}{\bibfnamefont{D.~L.} \bibnamefont{Bryce}} \bibnamefont{and}
  \bibinfo{author}{\bibfnamefont{R.~E.} \bibnamefont{Wasylishen}},
  \bibinfo{journal}{Acc. Chem. Res.} \textbf{\bibinfo{volume}{36}},
  \bibinfo{pages}{327} (\bibinfo{year}{2003}).

\bibitem[{\citenamefont{Mills et~al.}(1988)\citenamefont{Mills, Cvita\v{s},
  Homann, Kallay, and Kuchitsu}}]{Mills:1988}
\bibinfo{author}{\bibfnamefont{I.}~\bibnamefont{Mills}},
  \bibinfo{author}{\bibfnamefont{T.}~\bibnamefont{Cvita\v{s}}},
  \bibinfo{author}{\bibfnamefont{K.}~\bibnamefont{Homann}},
  \bibinfo{author}{\bibfnamefont{N.}~\bibnamefont{Kallay}}, \bibnamefont{and}
  \bibinfo{author}{\bibfnamefont{K.}~\bibnamefont{Kuchitsu}},
  \emph{\bibinfo{title}{Quantities, Units and Symbols in Physical Chemistry}}
  (\bibinfo{publisher}{Blackwell}, \bibinfo{address}{Oxford},
  \bibinfo{year}{1988}).

\bibitem[{\citenamefont{Kotochigova et~al.}(2003)\citenamefont{Kotochigova,
  Julienne, and Tiesinga}}]{Kotochigova:2003}
\bibinfo{author}{\bibfnamefont{S.}~\bibnamefont{Kotochigova}},
  \bibinfo{author}{\bibfnamefont{P.~S.} \bibnamefont{Julienne}},
  \bibnamefont{and} \bibinfo{author}{\bibfnamefont{E.}~\bibnamefont{Tiesinga}},
  \bibinfo{journal}{Phys. Rev. A} \textbf{\bibinfo{volume}{68}},
  \bibinfo{pages}{022501} (\bibinfo{year}{2003}).

\bibitem[{\citenamefont{Kotochigova and Tiesinga}(2005)}]{Kotochigova:2005}
\bibinfo{author}{\bibfnamefont{S.}~\bibnamefont{Kotochigova}} \bibnamefont{and}
  \bibinfo{author}{\bibfnamefont{E.}~\bibnamefont{Tiesinga}},
  \bibinfo{journal}{J. Chem. Phys.} \textbf{\bibinfo{volume}{123}},
  \bibinfo{pages}{174304} (\bibinfo{year}{2005}).

\bibitem[{\citenamefont{Van~Esbroeck et~al.}(1985)\citenamefont{Van~Esbroeck,
  McLean, Gaily, Holt, and Rosner}}]{Esbroeck:1985}
\bibinfo{author}{\bibfnamefont{P.~E.} \bibnamefont{Van~Esbroeck}},
  \bibinfo{author}{\bibfnamefont{R.~A.} \bibnamefont{McLean}},
  \bibinfo{author}{\bibfnamefont{T.~D.} \bibnamefont{Gaily}},
  \bibinfo{author}{\bibfnamefont{R.~A.} \bibnamefont{Holt}}, \bibnamefont{and}
  \bibinfo{author}{\bibfnamefont{S.~D.} \bibnamefont{Rosner}},
  \bibinfo{journal}{Phys. Rev. A} \textbf{\bibinfo{volume}{32}},
  \bibinfo{pages}{2595} (\bibinfo{year}{1985}).

\bibitem[{\citenamefont{te~Velde et~al.}(2001)\citenamefont{te~Velde,
  Bickelhaupt, van Gisbergen, Fonseca~Guerra, Baerends, Snijders, and
  Ziegler}}]{ADF1}
\bibinfo{author}{\bibfnamefont{G.}~\bibnamefont{te~Velde}},
  \bibinfo{author}{\bibfnamefont{F.~M.} \bibnamefont{Bickelhaupt}},
  \bibinfo{author}{\bibfnamefont{S.~J.~A.} \bibnamefont{van Gisbergen}},
  \bibinfo{author}{\bibfnamefont{C.}~\bibnamefont{Fonseca~Guerra}},
  \bibinfo{author}{\bibfnamefont{E.~J.} \bibnamefont{Baerends}},
  \bibinfo{author}{\bibfnamefont{J.~G.} \bibnamefont{Snijders}},
  \bibnamefont{and} \bibinfo{author}{\bibfnamefont{T.}~\bibnamefont{Ziegler}},
  \bibinfo{journal}{J. Comput. Chem.} \textbf{\bibinfo{volume}{22}},
  \bibinfo{pages}{931} (\bibinfo{year}{2001}).

\bibitem[{ADF(2007)}]{ADF3}
\emph{\bibinfo{title}{{ADF2007.01}}},
  \bibinfo{howpublished}{http://www.scm.com} (\bibinfo{year}{2007}),
  \bibinfo{note}{{SCM}, {T}heoretical {C}hemistry{,} {V}rije {U}niversiteit{,}
  {A}msterdam{,} {T}he {N}etherlands}.

\bibitem[{Dal(2005)}]{Dalton}
\emph{\bibinfo{title}{{DALTON}, a molecular electronic structure program,
  {R}elease 2.0}},
  \bibinfo{howpublished}{http://www.kjemi.uio.no/software/dalton/dalton.html}
  (\bibinfo{year}{2005}).

\bibitem[{\citenamefont{Logan et~al.}(1952)\citenamefont{Logan, Cot$\acute{\rm
  e}$, and Kusch}}]{Logan:1952}
\bibinfo{author}{\bibfnamefont{R.~A.} \bibnamefont{Logan}},
  \bibinfo{author}{\bibfnamefont{R.~E.} \bibnamefont{Cot$\acute{\rm e}$}},
  \bibnamefont{and} \bibinfo{author}{\bibfnamefont{P.}~\bibnamefont{Kusch}},
  \bibinfo{journal}{Phys. Rev.} \textbf{\bibinfo{volume}{86}},
  \bibinfo{pages}{280} (\bibinfo{year}{1952}).

\bibitem[{\citenamefont{Bonczyk and Hughes}(1967)}]{Bonczyk:1967}
\bibinfo{author}{\bibfnamefont{P.~A.} \bibnamefont{Bonczyk}} \bibnamefont{and}
  \bibinfo{author}{\bibfnamefont{V.~W.} \bibnamefont{Hughes}},
  \bibinfo{journal}{Phys. Rev.} \textbf{\bibinfo{volume}{161}},
  \bibinfo{pages}{15} (\bibinfo{year}{1967}).

\bibitem[{\citenamefont{Dagdigian and Wharton}(1972)}]{Dagdigian:1972}
\bibinfo{author}{\bibfnamefont{P.~J.} \bibnamefont{Dagdigian}}
  \bibnamefont{and} \bibinfo{author}{\bibfnamefont{L.}~\bibnamefont{Wharton}},
  \bibinfo{journal}{J. Chem. Phys.} \textbf{\bibinfo{volume}{57}},
  \bibinfo{pages}{1487} (\bibinfo{year}{1972}).

\bibitem[{\citenamefont{Cederberg
  et~al.}(2006{\natexlab{a}})\citenamefont{Cederberg, Frodermann, Tollerud,
  Huber, Bongard, Randolph, and Nitz}}]{Cederberg:2006}
\bibinfo{author}{\bibfnamefont{J.}~\bibnamefont{Cederberg}},
  \bibinfo{author}{\bibfnamefont{E.}~\bibnamefont{Frodermann}},
  \bibinfo{author}{\bibfnamefont{H.}~\bibnamefont{Tollerud}},
  \bibinfo{author}{\bibfnamefont{K.}~\bibnamefont{Huber}},
  \bibinfo{author}{\bibfnamefont{M.}~\bibnamefont{Bongard}},
  \bibinfo{author}{\bibfnamefont{J.}~\bibnamefont{Randolph}}, \bibnamefont{and}
  \bibinfo{author}{\bibfnamefont{D.}~\bibnamefont{Nitz}}, \bibinfo{journal}{J.
  Chem. Phys.} \textbf{\bibinfo{volume}{124}}, \bibinfo{pages}{244304}
  (\bibinfo{year}{2006}{\natexlab{a}}).

\bibitem[{\citenamefont{Cederberg
  et~al.}(2006{\natexlab{b}})\citenamefont{Cederberg, Fortman, Porter, Etten,
  Feig, Bongard, and Langer}}]{Cederberg:2006a}
\bibinfo{author}{\bibfnamefont{J.}~\bibnamefont{Cederberg}},
  \bibinfo{author}{\bibfnamefont{S.}~\bibnamefont{Fortman}},
  \bibinfo{author}{\bibfnamefont{B.}~\bibnamefont{Porter}},
  \bibinfo{author}{\bibfnamefont{M.}~\bibnamefont{Etten}},
  \bibinfo{author}{\bibfnamefont{M.}~\bibnamefont{Feig}},
  \bibinfo{author}{\bibfnamefont{M.}~\bibnamefont{Bongard}}, \bibnamefont{and}
  \bibinfo{author}{\bibfnamefont{L.}~\bibnamefont{Langer}},
  \bibinfo{journal}{J. Chem. Phys.} \textbf{\bibinfo{volume}{124}},
  \bibinfo{pages}{244305} (\bibinfo{year}{2006}{\natexlab{b}}).

\bibitem[{\citenamefont{Tiemann et~al.}(1977)\citenamefont{Tiemann, Holzer, and
  Hoeft}}]{Tiemann:1977}
\bibinfo{author}{\bibfnamefont{E.}~\bibnamefont{Tiemann}},
  \bibinfo{author}{\bibfnamefont{B.}~\bibnamefont{Holzer}}, \bibnamefont{and}
  \bibinfo{author}{\bibfnamefont{J.}~\bibnamefont{Hoeft}}, \bibinfo{journal}{Z.
  Naturforsch. A} \textbf{\bibinfo{volume}{32}}, \bibinfo{pages}{123}
  (\bibinfo{year}{1977}).

\bibitem[{\citenamefont{Tiemann et~al.}(1976)\citenamefont{Tiemann, Holzer, and
  Hoeft}}]{Tiemann:1976}
\bibinfo{author}{\bibfnamefont{E.}~\bibnamefont{Tiemann}},
  \bibinfo{author}{\bibfnamefont{B.}~\bibnamefont{Holzer}}, \bibnamefont{and}
  \bibinfo{author}{\bibfnamefont{J.}~\bibnamefont{Hoeft}}, \bibinfo{journal}{Z.
  Naturforsch. A} \textbf{\bibinfo{volume}{31}}, \bibinfo{pages}{236}
  (\bibinfo{year}{1976}).

\bibitem[{\citenamefont{Cederberg et~al.}(1999)\citenamefont{Cederberg, Ward,
  McAlister, Hilk, Beall, and Olson}}]{Cederberg:1999}
\bibinfo{author}{\bibfnamefont{J.}~\bibnamefont{Cederberg}},
  \bibinfo{author}{\bibfnamefont{J.}~\bibnamefont{Ward}},
  \bibinfo{author}{\bibfnamefont{G.}~\bibnamefont{McAlister}},
  \bibinfo{author}{\bibfnamefont{G.}~\bibnamefont{Hilk}},
  \bibinfo{author}{\bibfnamefont{E.}~\bibnamefont{Beall}}, \bibnamefont{and}
  \bibinfo{author}{\bibfnamefont{D.}~\bibnamefont{Olson}}, \bibinfo{journal}{J.
  Chem. Phys.} \textbf{\bibinfo{volume}{111}}, \bibinfo{pages}{8396}
  (\bibinfo{year}{1999}).

\bibitem[{\citenamefont{Hoeft et~al.}(1972)\citenamefont{Hoeft, Tiemann, and
  Torring}}]{Hoeft:1972}
\bibinfo{author}{\bibfnamefont{J.}~\bibnamefont{Hoeft}},
  \bibinfo{author}{\bibfnamefont{E.}~\bibnamefont{Tiemann}}, \bibnamefont{and}
  \bibinfo{author}{\bibfnamefont{T.}~\bibnamefont{Torring}},
  \bibinfo{journal}{Z. Naturforsch. A} \textbf{\bibinfo{volume}{27}},
  \bibinfo{pages}{1516} (\bibinfo{year}{1972}).

\bibitem[{\citenamefont{Brooks et~al.}(1963)\citenamefont{Brooks, Anderson, and
  Ramsey}}]{Brooks:1963}
\bibinfo{author}{\bibfnamefont{R.~A.} \bibnamefont{Brooks}},
  \bibinfo{author}{\bibfnamefont{C.~H.} \bibnamefont{Anderson}},
  \bibnamefont{and} \bibinfo{author}{\bibfnamefont{N.~F.}
  \bibnamefont{Ramsey}}, \bibinfo{journal}{Phys. Rev. Letters}
  \textbf{\bibinfo{volume}{10}}, \bibinfo{pages}{441} (\bibinfo{year}{1963}).

\bibitem[{\citenamefont{Brooks et~al.}(1972)\citenamefont{Brooks, Anderson, and
  Ramsey}}]{Brooks:1972}
\bibinfo{author}{\bibfnamefont{R.~A.} \bibnamefont{Brooks}},
  \bibinfo{author}{\bibfnamefont{C.~H.} \bibnamefont{Anderson}},
  \bibnamefont{and} \bibinfo{author}{\bibfnamefont{N.~F.}
  \bibnamefont{Ramsey}}, \bibinfo{journal}{J. Chem. Phys.}
  \textbf{\bibinfo{volume}{56}}, \bibinfo{pages}{5193} (\bibinfo{year}{1972}).

\bibitem[{\citenamefont{van Lenthe et~al.}(1993)\citenamefont{van Lenthe,
  Baerends, and Snijders}}]{Lenthe:1993}
\bibinfo{author}{\bibfnamefont{E.}~\bibnamefont{van Lenthe}},
  \bibinfo{author}{\bibfnamefont{E.~J.} \bibnamefont{Baerends}},
  \bibnamefont{and} \bibinfo{author}{\bibfnamefont{J.~G.}
  \bibnamefont{Snijders}}, \bibinfo{journal}{J. Chem. Phys.}
  \textbf{\bibinfo{volume}{99}}, \bibinfo{pages}{4597} (\bibinfo{year}{1993}).

\bibitem[{\citenamefont{van Lenthe et~al.}(1994)\citenamefont{van Lenthe,
  Baerends, and Snijders}}]{Lenthe:1994}
\bibinfo{author}{\bibfnamefont{E.}~\bibnamefont{van Lenthe}},
  \bibinfo{author}{\bibfnamefont{E.~J.} \bibnamefont{Baerends}},
  \bibnamefont{and} \bibinfo{author}{\bibfnamefont{J.~G.}
  \bibnamefont{Snijders}}, \bibinfo{journal}{J. Chem. Phys.}
  \textbf{\bibinfo{volume}{101}}, \bibinfo{pages}{9783} (\bibinfo{year}{1994}).

\bibitem[{\citenamefont{van Lenthe et~al.}(1999)\citenamefont{van Lenthe,
  Baerends, and Snijders}}]{Lenthe:1999}
\bibinfo{author}{\bibfnamefont{E.}~\bibnamefont{van Lenthe}},
  \bibinfo{author}{\bibfnamefont{E.~J.} \bibnamefont{Baerends}},
  \bibnamefont{and} \bibinfo{author}{\bibfnamefont{J.~G.}
  \bibnamefont{Snijders}}, \bibinfo{journal}{J. Chem. Phys.}
  \textbf{\bibinfo{volume}{110}}, \bibinfo{pages}{8943} (\bibinfo{year}{1999}).

\bibitem[{\citenamefont{Fedotov and Malkin}(1996)}]{Fedotov:1996}
\bibinfo{author}{\bibfnamefont{O.~L.} \bibnamefont{Fedotov},
  \bibfnamefont{M.~A.~Malkina}} \bibnamefont{and}
  \bibinfo{author}{\bibfnamefont{V.~G.} \bibnamefont{Malkin}},
  \bibinfo{journal}{Chem. Phys. Lett.} \textbf{\bibinfo{volume}{258}},
  \bibinfo{pages}{330} (\bibinfo{year}{1996}).

\bibitem[{\citenamefont{Bailey}(1998{\natexlab{a}})}]{Bailey:1998}
\bibinfo{author}{\bibfnamefont{W.~C.} \bibnamefont{Bailey}},
  \bibinfo{journal}{J. Mol. Spectrosc.} \textbf{\bibinfo{volume}{190}},
  \bibinfo{pages}{318} (\bibinfo{year}{1998}{\natexlab{a}}).

\bibitem[{\citenamefont{Bailey}(1998{\natexlab{b}})}]{Bailey:1998a}
\bibinfo{author}{\bibfnamefont{W.~C.} \bibnamefont{Bailey}},
  \bibinfo{journal}{Chem. Phys. Lett.} \textbf{\bibinfo{volume}{292}},
  \bibinfo{pages}{71} (\bibinfo{year}{1998}{\natexlab{b}}).

\bibitem[{\citenamefont{Bailey}(2000)}]{Bailey:2000}
\bibinfo{author}{\bibfnamefont{W.~C.} \bibnamefont{Bailey}},
  \bibinfo{journal}{Chem. Phys.} \textbf{\bibinfo{volume}{252}},
  \bibinfo{pages}{57} (\bibinfo{year}{2000}).

\bibitem[{\citenamefont{van Lenthe and Baerends}(2000)}]{Lenthe:2000}
\bibinfo{author}{\bibfnamefont{E.}~\bibnamefont{van Lenthe}} \bibnamefont{and}
  \bibinfo{author}{\bibfnamefont{E.~J.} \bibnamefont{Baerends}},
  \bibinfo{journal}{J. Chem. Phys.} \textbf{\bibinfo{volume}{112}},
  \bibinfo{pages}{8279} (\bibinfo{year}{2000}).

\bibitem[{\citenamefont{Hung and Schurko}(2003)}]{Hung:2003}
\bibinfo{author}{\bibfnamefont{I.}~\bibnamefont{Hung}} \bibnamefont{and}
  \bibinfo{author}{\bibfnamefont{R.~W.} \bibnamefont{Schurko}},
  \bibinfo{journal}{Solid State Nucl. Magn. Reson.}
  \textbf{\bibinfo{volume}{24}}, \bibinfo{pages}{78} (\bibinfo{year}{2003}).

\bibitem[{\citenamefont{Palmer and Nelson}(2007)}]{Palmer:2007}
\bibinfo{author}{\bibfnamefont{M.~H.} \bibnamefont{Palmer}} \bibnamefont{and}
  \bibinfo{author}{\bibfnamefont{A.~D.} \bibnamefont{Nelson}},
  \bibinfo{journal}{J. Mol. Struct.} \textbf{\bibinfo{volume}{828}},
  \bibinfo{pages}{91} (\bibinfo{year}{2007}).

\bibitem[{\citenamefont{Bischoff et~al.}(2007)\citenamefont{Bischoff,
  H$\ddot{\rm u}$bner, Klopper, Schnelzer, Pilawa, Horvati$\acute{\rm c}$, and
  Berthier}}]{Bischoff:2007}
\bibinfo{author}{\bibfnamefont{F.~A.} \bibnamefont{Bischoff}},
  \bibinfo{author}{\bibfnamefont{O.}~\bibnamefont{H$\ddot{\rm u}$bner}},
  \bibinfo{author}{\bibfnamefont{W.}~\bibnamefont{Klopper}},
  \bibinfo{author}{\bibfnamefont{L.}~\bibnamefont{Schnelzer}},
  \bibinfo{author}{\bibfnamefont{B.}~\bibnamefont{Pilawa}},
  \bibinfo{author}{\bibfnamefont{M.}~\bibnamefont{Horvati$\acute{\rm c}$}},
  \bibnamefont{and} \bibinfo{author}{\bibfnamefont{C.}~\bibnamefont{Berthier}},
  \bibinfo{journal}{Eur. Phys. J. B} \textbf{\bibinfo{volume}{55}},
  \bibinfo{pages}{229} (\bibinfo{year}{2007}).

\bibitem[{\citenamefont{Behzadi et~al.}(2007)\citenamefont{Behzadi, Hadipour,
  and Mirzaei}}]{Behzadi:2007}
\bibinfo{author}{\bibfnamefont{H.}~\bibnamefont{Behzadi}},
  \bibinfo{author}{\bibfnamefont{N.~L.} \bibnamefont{Hadipour}},
  \bibnamefont{and} \bibinfo{author}{\bibfnamefont{M.}~\bibnamefont{Mirzaei}},
  \bibinfo{journal}{Biophysical Chemistry} \textbf{\bibinfo{volume}{125}},
  \bibinfo{pages}{179} (\bibinfo{year}{2007}).

\bibitem[{\citenamefont{Lee et~al.}(1988)\citenamefont{Lee, Yang, and
  G.}}]{lee:1988}
\bibinfo{author}{\bibfnamefont{C.}~\bibnamefont{Lee}},
  \bibinfo{author}{\bibfnamefont{W.}~\bibnamefont{Yang}}, \bibnamefont{and}
  \bibinfo{author}{\bibfnamefont{P.~R.} \bibnamefont{G.}},
  \bibinfo{journal}{Phys. Rev. B} \textbf{\bibinfo{volume}{37}},
  \bibinfo{pages}{785} (\bibinfo{year}{1988}).

\bibitem[{\citenamefont{Becke}(1993)}]{becke:1993}
\bibinfo{author}{\bibfnamefont{A.~D.} \bibnamefont{Becke}},
  \bibinfo{journal}{J. Chem. Phys.} \textbf{\bibinfo{volume}{98}},
  \bibinfo{pages}{5648} (\bibinfo{year}{1993}).

\bibitem[{\citenamefont{Keal and Tozer}(2003)}]{Keal:2003}
\bibinfo{author}{\bibfnamefont{T.~W.} \bibnamefont{Keal}} \bibnamefont{and}
  \bibinfo{author}{\bibfnamefont{D.~J.} \bibnamefont{Tozer}},
  \bibinfo{journal}{J. Chem. Phys.} \textbf{\bibinfo{volume}{119}},
  \bibinfo{pages}{3015} (\bibinfo{year}{2003}).

\bibitem[{\citenamefont{Keal et~al.}(2004)\citenamefont{Keal, Tozer, and
  Helgaker}}]{Keal:2004}
\bibinfo{author}{\bibfnamefont{T.~W.} \bibnamefont{Keal}},
  \bibinfo{author}{\bibfnamefont{D.~J.} \bibnamefont{Tozer}}, \bibnamefont{and}
  \bibinfo{author}{\bibfnamefont{T.}~\bibnamefont{Helgaker}},
  \bibinfo{journal}{Chem. Phys. Lett.} \textbf{\bibinfo{volume}{391}},
  \bibinfo{pages}{374} (\bibinfo{year}{2004}).

\bibitem[{\citenamefont{Becke}(1988)}]{becke:1988}
\bibinfo{author}{\bibfnamefont{A.~D.} \bibnamefont{Becke}},
  \bibinfo{journal}{Phys. Rev. A} \textbf{\bibinfo{volume}{38}},
  \bibinfo{pages}{3098} (\bibinfo{year}{1988}).

\bibitem[{\citenamefont{Vaara et~al.}(2002)\citenamefont{Vaara, Jokisaari,
  Wasylishen, and Bryce}}]{Vaara:2002}
\bibinfo{author}{\bibfnamefont{J.}~\bibnamefont{Vaara}},
  \bibinfo{author}{\bibfnamefont{J.}~\bibnamefont{Jokisaari}},
  \bibinfo{author}{\bibfnamefont{R.~E.} \bibnamefont{Wasylishen}},
  \bibnamefont{and} \bibinfo{author}{\bibfnamefont{D.~L.} \bibnamefont{Bryce}},
  \bibinfo{journal}{Prog. Nucl. Magn. Reson. Spectrosc.}
  \textbf{\bibinfo{volume}{41}}, \bibinfo{pages}{233} (\bibinfo{year}{2002}).

\bibitem[{\citenamefont{Perdew et~al.}(1996)\citenamefont{Perdew, Burke, and
  Ernzerhof}}]{perdew:1996}
\bibinfo{author}{\bibfnamefont{J.~P.} \bibnamefont{Perdew}},
  \bibinfo{author}{\bibfnamefont{K.}~\bibnamefont{Burke}}, \bibnamefont{and}
  \bibinfo{author}{\bibfnamefont{M.}~\bibnamefont{Ernzerhof}},
  \bibinfo{journal}{Phys. Rev. Lett.} \textbf{\bibinfo{volume}{77}},
  \bibinfo{pages}{3865} (\bibinfo{year}{1996}).

\bibitem[{\citenamefont{Flygare}(1964)}]{Flygare:1964}
\bibinfo{author}{\bibfnamefont{W.~H.} \bibnamefont{Flygare}},
  \bibinfo{journal}{J. Chem. Phys.} \textbf{\bibinfo{volume}{41}},
  \bibinfo{pages}{793} (\bibinfo{year}{1964}).

\bibitem[{\citenamefont{Gierke and Flygare}(1972)}]{Gierke:1972}
\bibinfo{author}{\bibfnamefont{T.~D.} \bibnamefont{Gierke}} \bibnamefont{and}
  \bibinfo{author}{\bibfnamefont{W.~H.} \bibnamefont{Flygare}},
  \bibinfo{journal}{J. Am. Chem. Soc.} \textbf{\bibinfo{volume}{94}},
  \bibinfo{pages}{7277} (\bibinfo{year}{1972}).

\bibitem[{\citenamefont{Wasylishen et~al.}(2000)\citenamefont{Wasylishen,
  Bryce, Evans, and Gerry}}]{Wasylishen:2000}
\bibinfo{author}{\bibfnamefont{R.~E.} \bibnamefont{Wasylishen}},
  \bibinfo{author}{\bibfnamefont{D.~L.} \bibnamefont{Bryce}},
  \bibinfo{author}{\bibfnamefont{C.~J.} \bibnamefont{Evans}}, \bibnamefont{and}
  \bibinfo{author}{\bibfnamefont{M.~C.~L.} \bibnamefont{Gerry}},
  \bibinfo{journal}{J. Mol. Spectrosc.} \textbf{\bibinfo{volume}{204}},
  \bibinfo{pages}{184} (\bibinfo{year}{2000}).

\bibitem[{\citenamefont{Cooke and Gerry}(2004)}]{Cooke:2004}
\bibinfo{author}{\bibfnamefont{S.~A.} \bibnamefont{Cooke}} \bibnamefont{and}
  \bibinfo{author}{\bibfnamefont{M.~C.~L.} \bibnamefont{Gerry}},
  \bibinfo{journal}{Phys. Chem. Chem. Phys.} \textbf{\bibinfo{volume}{6}},
  \bibinfo{pages}{4579} (\bibinfo{year}{2004}).

\bibitem[{\citenamefont{Huzinaga and Miguel}(1990)}]{Huzinaga:1990}
\bibinfo{author}{\bibfnamefont{S.}~\bibnamefont{Huzinaga}} \bibnamefont{and}
  \bibinfo{author}{\bibfnamefont{B.}~\bibnamefont{Miguel}},
  \bibinfo{journal}{Chem. Phys. Lett.} \textbf{\bibinfo{volume}{175}},
  \bibinfo{pages}{289} (\bibinfo{year}{1990}).

\bibitem[{\citenamefont{Huzinaga and Klobukowski}(1993)}]{Huzinaga:1993}
\bibinfo{author}{\bibfnamefont{S.}~\bibnamefont{Huzinaga}} \bibnamefont{and}
  \bibinfo{author}{\bibfnamefont{M.}~\bibnamefont{Klobukowski}},
  \bibinfo{journal}{Chem. Phys. Lett.} \textbf{\bibinfo{volume}{212}},
  \bibinfo{pages}{260} (\bibinfo{year}{1993}).

\bibitem[{\citenamefont{Wilson et~al.}(2005)\citenamefont{Wilson, Mohn, and
  Helgaker}}]{Wilson:2005}
\bibinfo{author}{\bibfnamefont{D.~J.~D.} \bibnamefont{Wilson}},
  \bibinfo{author}{\bibfnamefont{C.~E.} \bibnamefont{Mohn}}, \bibnamefont{and}
  \bibinfo{author}{\bibfnamefont{T.}~\bibnamefont{Helgaker}},
  \bibinfo{journal}{J. Chem. Theory Comput.} \textbf{\bibinfo{volume}{1}},
  \bibinfo{pages}{877} (\bibinfo{year}{2005}).

\bibitem[{\citenamefont{Enevoldsen et~al.}(2001)\citenamefont{Enevoldsen,
  Rasmussen, and Sauer}}]{Enevoldsen:2001}
\bibinfo{author}{\bibfnamefont{T.}~\bibnamefont{Enevoldsen}},
  \bibinfo{author}{\bibfnamefont{T.}~\bibnamefont{Rasmussen}},
  \bibnamefont{and} \bibinfo{author}{\bibfnamefont{S.~P.~A.}
  \bibnamefont{Sauer}}, \bibinfo{journal}{J. Chem. Phys.}
  \textbf{\bibinfo{volume}{114}}, \bibinfo{pages}{84} (\bibinfo{year}{2001}).

\bibitem[{\citenamefont{Ross et~al.}(1990)\citenamefont{Ross, Effantin, Crozet,
  and Boursey}}]{Ross:1990}
\bibinfo{author}{\bibfnamefont{A.~J.} \bibnamefont{Ross}},
  \bibinfo{author}{\bibfnamefont{C.}~\bibnamefont{Effantin}},
  \bibinfo{author}{\bibfnamefont{P.}~\bibnamefont{Crozet}}, \bibnamefont{and}
  \bibinfo{author}{\bibfnamefont{E.}~\bibnamefont{Boursey}},
  \bibinfo{journal}{J. Phys. B: At. Mol. Opt. Phys.}
  \textbf{\bibinfo{volume}{23}}, \bibinfo{pages}{L247} (\bibinfo{year}{1990}).

\bibitem[{\citenamefont{Kat$\hat{{\rm o}}$ and Kobayashi}(1983)}]{Kato:1983}
\bibinfo{author}{\bibfnamefont{H.}~\bibnamefont{Kat$\hat{{\rm o}}$}}
  \bibnamefont{and}
  \bibinfo{author}{\bibfnamefont{H.}~\bibnamefont{Kobayashi}},
  \bibinfo{journal}{J. Chem. Phys.} \textbf{\bibinfo{volume}{79}},
  \bibinfo{pages}{123} (\bibinfo{year}{1983}).

\bibitem[{\citenamefont{Lawrence et~al.}(1963)\citenamefont{Lawrence, Anderson,
  and Ramsey}}]{Lawrence:1963}
\bibinfo{author}{\bibfnamefont{T.~R.} \bibnamefont{Lawrence}},
  \bibinfo{author}{\bibfnamefont{C.~H.} \bibnamefont{Anderson}},
  \bibnamefont{and} \bibinfo{author}{\bibfnamefont{N.~F.}
  \bibnamefont{Ramsey}}, \bibinfo{journal}{Phys. Rev.}
  \textbf{\bibinfo{volume}{130}}, \bibinfo{pages}{1865} (\bibinfo{year}{1963}).

\bibitem[{\citenamefont{Zare}(1987)}]{Zare}
\bibinfo{author}{\bibfnamefont{R.~N.} \bibnamefont{Zare}},
  \emph{\bibinfo{title}{Angular Momentum}} (\bibinfo{publisher}{John Wiley \&
  Sons}, \bibinfo{year}{1987}).

\bibitem[{\citenamefont{Townes and Schawlow}(1975)}]{Townes}
\bibinfo{author}{\bibfnamefont{R.~N.} \bibnamefont{Townes}} \bibnamefont{and}
  \bibinfo{author}{\bibfnamefont{A.~L.} \bibnamefont{Schawlow}},
  \emph{\bibinfo{title}{Microwave Spectroscopy}} (\bibinfo{publisher}{Dover
  Publications, New York}, \bibinfo{year}{1975}).

\end{thebibliography}

\end{document}